\documentclass[12pt]{article}
\usepackage{amsmath}

\def\ex#1{\langle#1\rangle}
\def\W#1{W_{\text{#1}}}
\def\tr{\mathop{\text{tr}}}

\def\SU{SU}

\def\Nis{\mathcal{N}=}
\def\Pf{\mathop{\text{Pf}}}

\begin{document}
\thispagestyle{empty}
\hbox{}

\begin{flushright}
UT-02-62\\
hep-th/0211274
\end{flushright}

\bigskip\bigskip\bigskip
\bigskip\bigskip\bigskip

\begin{center}
\LARGE

Derivation of the linearity principle
of Intriligator-Leigh-Seiberg

\bigskip

{\Large
Yuji Tachikawa}\\

\normalsize
\bigskip

\textit{
Department of Physics, Faculty of Science, University of Tokyo,\\
Hongo 7-3-1, Bunkyo-ku, Tokyo 113-0033, Japan}

\bigskip

email: \texttt{yujitach@hep-th.phys.s.u-tokyo.ac.jp}
\end{center}

\textbf{Abstract:}
Utilizing the techniques recently developed
for $\Nis1$ super Yang-Mills theories
by Dijkgraaf, Vafa and collaborators, 
we derive the linearity principle of Intriligator, Leigh and Seiberg,
for the confinement phase of the theories
with semi-simple gauge groups and matters in a non-chiral representation
which satisfies a further technical assumption.

\newpage\setcounter{page}{1}
\section{Introduction and Results}
Recently Dijkgraaf and Vafa proposed
in \cite{Dijkgraaf:2002dh}
that the non-perturbative superpotential $\W{eff}$ of
$\Nis1$ super Yang-Mills theories is captured by
the perturbative calculation for the holomorphic matrix models.
This triggered an avalanche of works \cite{CHECKS} checking 
and extending the proposal, and now we have two different,
purely field-theoretic
derivations for it \cite{Dijkgraaf:2002xd} \cite{Cachazo:2002ry}.
Originally the proposal were made for matters in adjoint or bi-fundamental
representations, recent works \cite{Flavors} extends this method
to matters in the fundamental representation.

In the method of Dijkgraaf and Vafa,
one first introduces a bare superpotential $\W{tree}=\sum g_iO_i$
which leads the system to the confinement phase,
and then the effective superpotential for the gaugino condensate $S$
is calculated perturbatively.
Therefore, to check the proposal,
one has to calculate the effective superpotential
by some other methods. Usually this is done by combining two ingredients:
one is the exact results
such as the Affleck-Dine-Seiberg superpotential or the Seiberg-Witten curves
which describes the system without $\W{tree}$,
and the other is the linearity principle of Intriligator, Leigh and Seiberg
\cite{Intriligator:1994jr}\cite{Intriligator:1994uk}
which governs the reaction of the system
to the bare superpotential $\W{tree}$.
For example, Ferrari showed in \cite{Ferrari:2002jp} that 
for the maximally-confined phase of
the $\Nis1$ $U(N)$ super Yang-Mills with one adjoint,
the effective superpotential calculated by the Dijkgraaf-Vafa method
exactly matches that calculated by the linearity principle
combined with the Seiberg-Witten solution. This indicates that the linearity
principle can be derived generically from the Dijkgraaf-Vafa proposal.

Let us now briefly review the linearity principle.
It states that when the effective superpotential
$\W{eff}$ is written as a sum of two terms $\W{n.p.}+\W{tree}$,
the non-perturbatively generated part $\W{n.p.}$
is independent of $g_i$ and is a function of
$\Lambda$ and $\ex{O_i}$ only, where $\Lambda$ denotes
the dynamical scale of the theory and $\ex{O_i}$
are the vacuum expectation values of
gauge-invariant combinations of matter chiral superfields.
Hence $\W{eff}$ is \textit{linear} in $g_i$'s.
The authors of \cite{Intriligator:1994jr}\cite{Intriligator:1994uk}
checked this principle for several simple cases by symmetry and holomorphy,
but for matters in more complicated representation,
symmetry and holomorphy themselves are not strong enough to ensure 
the absence of further correction to $\W{eff}$.
In the paper,
they proposed that
this linearity holds in general,
by suitably defining the composite operators.
Anomalous global symmetries played a significant role in the analysis.
An example is the $R$ symmetry, which transforms matter fermions $\psi$ 
and gauginos $\lambda$ as
$\psi(x)\to e^{-i\varphi}\psi(x)$ and
$\lambda (x)\to e^{i\varphi}\lambda(x)$,
respectively.

Now that we have derivations for the Dijkgraaf-Vafa proposal,
we can turn the argument around and 
derive the linearity principle 
for a general class of theories,
using the ideas in \cite{Dijkgraaf:2002xd} and
\cite{Cachazo:2002ry}.
We show the linearity for theories which
satisfy the following three criteria:

First, we consider only the phase where the gauge group is completely Higgsed
or completely confined, so that the low energy gaugino condensate
for each group is characterized by a single superfield $S$.

Second, if the $R$ charge of $\Lambda$
of the theory is non-zero, 
we further impose the following restriction: 

\noindent\textbf{Restriction A}\quad
There are $r$ basic gauge invariants $F_1$,\ldots,$F_r$
such that any of the gauge invariants $O_i$ can be
written as a polynomial of them,
and that no dynamical constraints $P(\ex{F_1},\ex{F_2},\ldots,\ex{F_r})=0$
are present among $F_i$'s.

Third, we also take the gauge group to be simple
for the sake of brevity. The extension to general semi-simple groups
should be immediate. 

Please note that many theories satisfy these criteria.
As an example, let us recall the $Sp(N)$ super Yang-Mills
 with $2N_f$ fundamentals $Q_i$.
Basic gauge invariants are $T_{ij}=Q_iQ_j$.
For $N_f<N+1$, there are no constraints.
For $N_f=N+1$, there is a dynamical constraint $\Pf T_{ij}=\Lambda^{2N_f}$,
but in this case the $R$ charge of $\Lambda$ is zero.
For $N_f>N+1$, there are classical constraints analogous to the one above,
but they are all lifted dynamically.
Therefore they all meet the criteria.
The behavior of the $SU(N)$ super Yang-Mills 
with fundamentals is similar.

The derivation  is carried out in the following two steps:
In section 2, we show that 
$\W{n.p.}$ is equal to $(N_c-N_f)S$ where $N_c$ is the dual Coxeter number
of the gauge group and $N_f$ is the index of anomaly of
the representation of the matter fields. 
In section 3, under the restriction \textbf{A},
we derive that $S$ is
independent of the coupling constants $g_i$ in $\W{tree}$,
 when expressed as a function of 
$\Lambda$ and the basic gauge invariants $\ex{F_i}$.
Combining the results obtained in section 2 and 3,
the linearity principle follows for the theories considered in this paper.
Section 4 contains the conclusion.

We follow the conventions in \cite{Dijkgraaf:2002xd}.

\section{Determination of $\W{n.p.}$}
Let us consider an $\Nis1$ super Yang-Mills system with
a simple gauge group and 
matter fields in some non-chiral representation so that
one can use the Feynman rules which are spelled out in \cite{Dijkgraaf:2002xd}.

According to \cite{Dijkgraaf:2002xd},
with the bare superpotential $\W{tree}$ introduced,
the effective superpotential $\W{eff}$ for the gaugino condensate
is calculated by introducing the vector superfield as an external
background and integrating the matter superfields out.
We only use a few essential features, namely  that
i) the propagator for each superfield is the inverse of its mass,
ii) the vertices come from terms in $\W{tree}$ which is higher than
or equal to cubic,
and iii) each loop integral brings in two factors of $W_\alpha$ whose
lowest component is the gaugino. $S$ is found by extremizing $\W{eff}$.

Another important feature is that, for pure super Yang-Mills theories
with a simple gauge group, the chiral ring is generated
by $S=\tr W_\alpha W^\alpha/(32\pi^2)$, as noted in \cite{Cachazo:2002ry}.
Hence, after integrating out all the matter superfields,
any possible contraction of $2n$ of $W_\alpha$'s becomes $S^n$ times
some numerical constant. Therefore any $n$ loop diagram is accompanied by
the factor $S^n$.

Let us denote $\W{tree}=\sum m_iQ_i+\sum g_jO_j$,
where $m_i$ is the mass of the $i^{\text{th}}$ superfield,
$Q_i$ are the corresponding quadratic gauge invariants,
$O_j$ are gauge invariants which is cubic or higher in matter superfields,
and $g_j$ are coupling constants for them.
For the $\SU(N)$ super Yang-Mills with an adjoint matter $\Phi$, for example,
one can introduce $\tr \Phi^n$ and $(\tr \Phi^n)^m$ as gauge invariants.
We can also introduce baryonic invariants if one have
enough number of fundamental matters.
Let us also denote by $n_i$ the index\footnote{%
The index of anomaly $T(r)$ of the representation $r$ is defined by the formula
$\tr_r(t^at^b)=T(r)\delta^{ab}$, where $t^a$'s are the generators of the
gauge group normalized so that for the standard $SU(2)$ subgroup
they become one half of the Pauli sigma matrices.
} of the representation of the
$i^{\text{th}}$ matter superfield, plus that of its conjugate if
the representation is not real.
For example for the gauge group $SU(N)$, 
$n_i=N$ for an adjoint and
$n_i=1$ for a pair of the fundamental and the anti-fundamental.

Once one has $\W{eff}$, 
one can calculate
the vacuum expectation values for various operators $O_i$ by
just differentiating $\W{eff}$ with respect to $g_i$ because\begin{equation}
\ex{O_i}=\frac{\partial}{\partial g_i}
\W{eff}(S(\Lambda,g_j),g_j)
=\frac{\partial\W{eff}}{\partial g_i}+\frac{\partial S}{\partial g_i}
\frac{\partial\W{eff}}{\partial S}
=\frac{\partial\W{eff}}{\partial g_i}
\end{equation}
where we used the extremization condition $\partial\W{eff}/\partial S=0$.

Now recall the effective superpotential $\W{eff}$ is a sum of three terms:
\begin{equation}
\W{eff}=\W{VY}+\W{one loop}+\W{higher}
\end{equation}
where $\W{VY}$ is the Veneziano-Yankielowicz term
$N_cS(1-\log(S/\Lambda^3))$, $\W{one loop}=\sum n_i S\log m_i/\Lambda$
is the one loop contribution without any vertex insertion,
and $\W{higher}=\sum D$ is the sum of diagrams with some vertex insertions
(refer \cite{Tachikawa:2002wk} for a more detailed explanation).
The non-perturbative part of the superpotential $\W{n.p.}$ is defined as
\begin{equation}
\W{n.p.}=\W{eff}-\ex{\W{tree}}.
\end{equation}

Let us note that the number of vertices $V$, the number of propagators $E$, 
and the number of loops $L$ satisfies $V-E+L=1$ and that
$V$, $E$ and  $L$ of each diagram
can be counted by \begin{equation}
\sum g_j\frac\partial{\partial g_j},
\qquad
\sum m_i^{-1}\frac{\partial}{\partial m_i^{-1}}
=-\sum m_i\frac{\partial}{\partial m_i},
\quad \text{and} \quad
S\frac{\partial}{\partial S}
\end{equation}respectively. Hence, for each diagram $D$, the value of $D$ satisfies
(we denote a diagram and its value by the same letter)\begin{equation}
\left(1-S\frac{\partial}{\partial S}\right)D
=\left(\sum m_i\frac{\partial}{\partial m_i}
+\sum g_j\frac\partial{\partial g_j}\right)D,
\end{equation} but the right hand side is just the contribution of the diagram $D$
to $\ex{\W{tree}}$. This means that \begin{align}
\ex{\W{tree}}&=\left(\sum m_i\frac{\partial}{\partial m_i}
+\sum g_j\frac\partial{\partial g_j}\right)(\W{one loop}+\W{higher})\nonumber\\
&=\left(\sum m_i\frac{\partial\W{one loop}}{\partial m_i}\right)+
\left(1-S\frac{\partial}{\partial S}\right)\W{higher}.
\end{align}Combining this with the equation $\partial\W{eff}/\partial S=0$,
one can derive\begin{align}
\W{n.p.}&=\W{eff}-\ex{\W{tree}}\nonumber\\
&=\W{VY}+\left(1-\sum m_i\frac{\partial}{\partial m_i}\right)\W{one loop}
+S\frac{\partial}{\partial S}\W{higher}\nonumber\\
&=\left(1-S\frac{\partial}{\partial S}\right)\W{VY}+\left(1-S\frac{\partial}{\partial S}-\sum m_i\frac{\partial}{\partial m_i}\right)\W{one loop}\nonumber\\
&=(N_c-\sum n_i)S.
\end{align}
This is the result of this section.
As an simple application of this, one can see that there is no 
non-perturbative superpotential generated for the $\SU(N_c)$
super Yang-Mills with $N_f=N_c$ pairs of fundamental flavors.

\section{Independence of $S$ from coupling constants}
In the following,
we assume the theory under consideration satisfies
the restriction \textbf{A}.
We do not need to distinguish quadratic operators
and operators which is higher than quadratic, so we collectively denote them
as $O_i$ and write $\W{tree}=\sum\lambda_iO_i$. We also take $\W{eff}$ as
a function of $\Lambda$, $S$, and coupling constants $\lambda_i$.
As a convention, we take $O_i=F_i$ for $i=1,2,\ldots,r$.

In this section, we show that the gaugino condensate $S$ can be written as
a function of $\Lambda$ and $\ex{F_i}$
without explicit dependence on $\lambda_i$.
In other words, we show that as long as
$\ex{F_i}$'s are left  invariant, $S$ does not change
when $\lambda_i$'s  are varied.

To show the claim above, first let us recall
that $S$ is determined by extremizing $\W{eff}$,
so that\begin{equation}
-\frac{\partial^2\W{eff}}{\partial S\partial S}\delta S
=\sum_i \delta\lambda_i\frac{\partial^2\W{eff}}{\partial\lambda_i\partial S}.
\end{equation}(Note that $-{\partial^2\W{eff}}/\partial S^2$ contains a term $N_c/S$
coming from the Veneziano-Yankielowicz contribution, so generally non-zero.)

Second, the change in vacuum expectation values induced by the change in coupling constants is
\begin{equation}
\delta\ex{F_j}=\delta\frac{\partial\W{eff}}{\partial\lambda_j}
=\delta S\frac{\partial^2\W{eff}}{\partial S\partial\lambda_j}+
\sum_i\delta \lambda_i
\frac{\partial^2\W{eff}}{\partial\lambda_i\partial\lambda_j}
=\sum_i\delta \lambda_i G_{ij}
\end{equation}where\begin{equation}
G_{ij}=
\frac{\partial^2\W{eff}}{\partial\lambda_i\partial\lambda_j}
-\frac{\partial^2\W{eff}}{\partial\lambda_i\partial S}
\frac{\partial^2\W{eff}}{\partial S\partial\lambda_j}\Big/
\frac{\partial^2\W{eff}}{\partial S\partial S}.
\end{equation}

Please note that if some dynamical constraints emerge,
the rank of $G_{ij}$ is less than $r$, the number of basic
gauge invariants.
Therefore, the restriction \textbf{A}
ensures that the rank of $G_{ij}$ is equal to $r$.

Let us view $\ex{O_i}=\partial\W{eff}/\partial\lambda_i$
as the expectation values of $O_i$ in the static background of the
gaugino condensate $S$.
Then, using the factorization of gauge invariants
$\delta\ex{O_i}=\sum_k\ex{\partial O_i/ \partial F_k}\delta\ex{F_k}$,
one can rewrite $\delta S$ as\begin{equation}
-\frac{\partial^2\W{eff}}{\partial S\partial S}\delta S
=\sum_i \delta\lambda_i\frac{\partial^2\W{eff}}{\partial\lambda_i\partial S}
=\sum_i \delta\lambda_i\frac{\partial\ex{O_i}}{\partial S}
=\sum_{k=1}^r\left(
\sum_i\delta\lambda_i\ex{\frac{\partial O_i}{\partial F_k}}\right)
\frac{\partial\ex{F_k}}{\partial S}
\end{equation}and $\delta\ex{F_j}$ as\begin{align}
\delta\ex{F_j}&=\sum_i\delta\lambda_i
\left(\frac{\partial \ex{O_i}}{\partial\lambda_j}
-\frac{\partial \ex{O_i}}{\partial S}
\frac{\partial^2\W{eff}}{\partial S\partial\lambda_j}\Big/
\frac{\partial^2\W{eff}}{\partial S\partial S}\right)\nonumber\\
&=\sum_{k=1}^r
\left(\sum_i\delta\lambda_i\ex{\frac{\partial{O_i}}{\partial{F_k}}}\right)
\left(\frac{\partial \ex{F_k}}{\partial\lambda_j}
-\frac{\partial \ex{F_k}}{\partial S}
\frac{\partial^2\W{eff}}{\partial S\partial\lambda_j}\Big/
\frac{\partial^2\W{eff}}{\partial S\partial S}\right)\nonumber\\
&=\sum_{k=1}^r
\left(\sum_i\delta\lambda_i\ex{\frac{\partial{O_i}}{\partial{F_k}}}\right)
G_{kj}.\label{QQQ}
\end{align} 
As already noted, $G_{kj}$ is invertible
if the indices $k$ and $j$ are restricted to the range $1,2,\ldots, r$.
Thus, by multiplying (\ref{QQQ}) by $G^{-1}{}_{jk}$, one obtains
\begin{equation}
\sum_i\delta\lambda_i\ex{\frac{\partial{O_i}}{\partial{F_k}}}=0
\end{equation}for $k=1,2,\ldots, r$, so that $\delta S=0$  follows.
This is the result of this section.

\section{Conclusion}
Let us combine the results obtained so far.
First, if the $R$ charge of $\Lambda$ is zero, $N_c-N_f$ is also zero.
Hence, the non-perturbatively generated superpotential satisfies
$\W{eff}=(N_c-N_f)S=0$ by the result of section 2.
Second, if the $R$ charge of $\Lambda$ is non-vanishing,
we  can use the result of section 3 because we assume
the theory satisfies the restriction \textbf{A}.
Hence $\W{eff}=(N_c-N_f)S$ 
does not explicitly depend on the coupling constants, and 
 is a function of $\Lambda$ and $\ex{F_i}$. 
This is what we wanted to derive.

The extension to a wider class of theories 
will be interesting and worth studying.
Extending the analysis to the chiral matter contents seems to be much harder,
because recent developments are mainly focused on theories with
matters in a non-chiral  representation.

\paragraph{Acknowledgments}
The author would like to thank T.~Eguchi, 
F.~Koyama,  Y.~Naka\-yama and R.~Nobuyama for  very helpful discussions.


\begin{thebibliography}{99}



\bibitem{Dijkgraaf:2002dh}
R.~Dijkgraaf and C.~Vafa,
arXiv:hep-th/0208048.


\bibitem{CHECKS}
L.~Chekhov and A.~Mironov,
arXiv:hep-th/0209085.

N.~Dorey, T.~J.~Hollowood, S.~P.~Kumar and A.~Sinkovics,
arXiv:hep-th/0209089.

N.~Dorey, T.~J.~Hollowood, S.~P.~Kumar and A.~Sinkovics,
arXiv:hep-th/0209099.




T.~J.~Hollowood and T.~Kingaby,
arXiv:hep-th/0210096.

H.~Fuji and Y.~Ookouchi,
arXiv:hep-th/0210148.

D.~Berenstein,
arXiv:hep-th/0210183.



R.~Dijkgraaf, S.~Gukov, V.~A.~Kazakov and C.~Vafa,
arXiv:hep-th/0210238.

N.~Dorey, T.~J.~Hollowood and S.~P.~Kumar,
arXiv:hep-th/0210239.

A.~Gorsky,
arXiv:hep-th/0210281.





F.~Ferrari,
arXiv:hep-th/0211069.




R.~Gopakumar,
arXiv:hep-th/0211100.

S.~G.~Naculich, H.~J.~Schnitzer and N.~Wyllard,
arXiv:hep-th/0211123.



R.~Dijkgraaf, A.~Neitzke and C.~Vafa,
arXiv:hep-th/0211194.

A.~Klemm, M.~Marino and S.~Theisen,
arXiv:hep-th/0211216.

H.~Ita, H.~Nieder and Y.~Oz,
arXiv:hep-th/0211261.



\bibitem{Dijkgraaf:2002xd}
R.~Dijkgraaf, M.~T.~Grisaru, C.~S.~Lam, C.~Vafa and D.~Zanon,
arXiv:hep-th/0211017.

\bibitem{Cachazo:2002ry}
F.~Cachazo, M.~Douglas, N.~Seiberg, and E.~Witten,
arXiv:hep-th/0211170.


\bibitem{Flavors}

R.~Argurio, V.~L.~Campos, G.~Ferretti and R.~Heise,
arXiv:hep-th/0210291.

J.~McGreevy,
arXiv:hep-th/0211009.

H.~Suzuki,
arXiv:hep-th/0211052.

I.~Bena and R.~Roiban,
arXiv:hep-th/0211075.

Y.~Demasure and R.~A.~Janik,
arXiv:hep-th/0211082.

B.~Feng,
arXiv:hep-th/0211202.

B.~Feng and Y.~H.~He,
arXiv:hep-th/0211234.

R.~Argurio, V.~L.~Campos, G.~Ferretti and R.~Heise,
arXiv:hep-th/0211249.

S.~Naculich, H.~Schnitzer and N.~Wyllard,
arXiv:hep-th/0211254.


\bibitem{Intriligator:1994jr}
K.~A.~Intriligator, R.~G.~Leigh and N.~Seiberg,
Phys.\ Rev.\ D {\bf 50}, 1092 (1994)
[arXiv:hep-th/9403198].

\bibitem{Intriligator:1994uk}
K.~A.~Intriligator,
Phys.\ Lett.\ B {\bf 336}, 409 (1994)
[arXiv:hep-th/9407106].

\bibitem{Ferrari:2002jp}
F.~Ferrari,
arXiv:hep-th/0210135.

\bibitem{Tachikawa:2002wk}
Y.~Tachikawa,
arXiv:hep-th/0211189.



\end{thebibliography}
\end{document}